\documentclass[aps,prl,superscriptaddress,showpacs,floatfix,amsmath,amssymb,twocolumn]{revtex4}
\usepackage{epsfig}
\usepackage{bm}
\usepackage{hyperref}
\usepackage{balance}
\usepackage{colordvi}\usepackage[usenames]{color}
\usepackage[T1]{fontenc}
\def\etal{{\it et al.}}
\newcommand{\header}[1]{{\em#1.---}}

\begin{document}
\title{{\boldmath{}Extending the Southern Shore of the Island of Inversion to $^{28}$F}}

\author{A.~Revel} 
\affiliation{Grand Acc\'el\'erateur National d'Ions Lourds (GANIL),
CEA/DRF-CNRS/IN2P3, Bvd Henri Becquerel, 14076 Caen, France}
\author{O.~Sorlin} 
\affiliation{Grand Acc\'el\'erateur National d'Ions Lourds (GANIL),
  CEA/DRF-CNRS/IN2P3, Bvd Henri Becquerel, 14076 Caen, France}
\author{F.M.~Marqu\'es}
\affiliation{LPC Caen,
ENSICAEN, Universit\'e de Caen, CNRS/IN2P3, F-14050 CAEN Cedex, France}
\author{Y.~Kondo}
\affiliation{Department of Physics, Tokyo Institute of Technology, 2-12-1 O-Okayama, Meguro, Tokyo 152-8551, Japan}
\author{J.~Kahlbow}
\affiliation{Institut f\"ur Kernphysik, Technische Universit\"at Darmstadt, 64289 Darmstadt, Germany}
\affiliation{RIKEN Nishina Center, Hirosawa 2-1, Wako, Saitama 351-0198, Japan}
\author{T.~Nakamura}
\affiliation{Department of Physics, Tokyo Institute of Technology, 2-12-1 O-Okayama, Meguro, Tokyo 152-8551, Japan}
\author{N.A.~Orr}
\affiliation{LPC Caen,
ENSICAEN, Universit\'e de Caen, CNRS/IN2P3, F-14050 CAEN Cedex, France}
\author{F.~Nowacki}
\affiliation{Universit\'e de Strasbourg, IPHC, 23 rue de Loess 67037 Strasbourg, France}
\affiliation{CNRS, UMR7178, 67037 Strasbourg, France}
\author{J.A.~Tostevin}
\affiliation{Department of Physics, University of Surrey, Guildford, Surrey GU2 7XH, United Kingdom}
\author{C.X.~Yuan}
\affiliation{Sino-French Institute of Nuclear Engineering and Technology, Sun Yat-Sen University, Zhuhai 519082, China}
\author{N.L.~Achouri}
\affiliation{LPC Caen,
  ENSICAEN, Universit\'e de Caen, CNRS/IN2P3, F-14050 CAEN Cedex, France}
\author{H.~Al~Falou}
\affiliation{Lebanese University, Beirut, Lebanon}
\author{L.~Atar}
\affiliation{Institut f\"ur Kernphysik, Technische Universit\"at Darmstadt, 64289 Darmstadt, Germany}
\author{T.~Aumann}
\affiliation{Institut f\"ur Kernphysik, Technische Universit\"at Darmstadt, 64289 Darmstadt, Germany}
\affiliation{GSI Helmholtzzentrum f\"ur Schwerionenforschung, 64291 Darmstadt, Germany}
\author{H.~Baba}
\affiliation{RIKEN Nishina Center, Hirosawa 2-1, Wako, Saitama 351-0198, Japan}
\author{K.~Boretzky}
\affiliation{GSI Helmholtzzentrum f\"ur Schwerionenforschung, 64291 Darmstadt, Germany}
\author{C.~Caesar}
\affiliation{Institut f\"ur Kernphysik, Technische Universit\"at Darmstadt, 64289 Darmstadt, Germany}
\affiliation{GSI Helmholtzzentrum f\"ur Schwerionenforschung, 64291 Darmstadt, Germany}
\author{D.~Calvet}
\affiliation{Irfu, CEA, Universit\'e Paris-Saclay, 91191 Gif-sur-Yvette, France}
\author{H.~Chae}
\affiliation{IBS, 55, Expo-ro, Yuseong-gu, Daejeon, Korea, 34126}
\author{N.~Chiga}
\affiliation{RIKEN Nishina Center, Hirosawa 2-1, Wako, Saitama 351-0198, Japan}
\author{A.~Corsi}
\affiliation{Irfu, CEA, Universit\'e Paris-Saclay, 91191 Gif-sur-Yvette, France}
\author{H.~L.~Crawford}
\affiliation{Nuclear Science Division, Lawrence Berkeley National Laboratory, Berkeley, CA 94720, USA}
\author{F.~Delaunay}
\affiliation{LPC Caen,
  ENSICAEN, Universit\'e de Caen, CNRS/IN2P3, F-14050 CAEN Cedex, France}
\author{A.~Delbart}
\affiliation{Irfu, CEA, Universit\'e Paris-Saclay, 91191 Gif-sur-Yvette, France}
\author{Q.~Deshayes}
\affiliation{LPC Caen,
  ENSICAEN, Universit\'e de Caen, CNRS/IN2P3, F-14050 CAEN Cedex, France}
\author{Z.~Dombr\'adi}
\affiliation{Institute of Nuclear Research, Atomki, 4001 Debrecen, Hungary}
\author{C.~A.~Douma}
\affiliation{KVI-CART, University of Groningen, Zernikelaan 25, 9747 AA Groningen, The Netherlands}
\author{Z.~Elekes}
\affiliation{Institute of Nuclear Research, Atomki, 4001 Debrecen, Hungary}
\author{P.~Fallon}
\affiliation{Nuclear Science Division, Lawrence Berkeley National Laboratory, Berkeley, CA 94720, USA}
\author{I.~Ga\v{s}pari\'c}
\affiliation{Ru{\dj}er Bo\v{s}kovi\'c Institute, HR-10002 Zagreb, Croatia}
\affiliation{RIKEN Nishina Center, Hirosawa 2-1, Wako, Saitama 351-0198, Japan}
\author{J.-M.~Gheller}
\affiliation{Irfu, CEA, Universit\'e Paris-Saclay, 91191 Gif-sur-Yvette, France}
\author{J.~Gibelin}
\affiliation{LPC Caen,
  ENSICAEN, Universit\'e de Caen, CNRS/IN2P3, F-14050 CAEN Cedex, France}
\author{A.~Gillibert}
\affiliation{Irfu, CEA, Universit\'e Paris-Saclay, 91191 Gif-sur-Yvette, France}
\author{M.~N.~Harakeh}
\affiliation{GSI Helmholtzzentrum f\"ur Schwerionenforschung, 64291 Darmstadt, Germany}
\affiliation{KVI-CART, University of Groningen, Zernikelaan 25, 9747 AA Groningen, The Netherlands}
\author{W.~He}
\affiliation{RIKEN Nishina Center, Hirosawa 2-1, Wako, Saitama 351-0198, Japan}
\author{A.~Hirayama}
\affiliation{Department of Physics, Tokyo Institute of Technology, 2-12-1 O-Okayama, Meguro, Tokyo 152-8551, Japan}
\author{C.R.~Hoffman}
\affiliation{Physics Division, Argonne National Laboratory, Argonne, Illinois 60439, USA}
\author{M.~Holl}
\affiliation{GSI Helmholtzzentrum f\"ur Schwerionenforschung, 64291 Darmstadt, Germany}
\author{A.~Horvat}
\affiliation{GSI Helmholtzzentrum f\"ur Schwerionenforschung, 64291 Darmstadt, Germany}
\author{\'A.~Horv\'ath}
\affiliation{E\"otv\"os Lor\'and University, P\'azm\'any P\'eter S\'et\'any 1/A, H-1117 Budapest, Hungary}
\author{J.W.~Hwang}
\affiliation{Department of Physics and Astronomy, Seoul National University, 1 Gwanak-ro, Gwanak-gu, Seoul 08826, Republic of Korea}
\author{T.~Isobe}
\affiliation{RIKEN Nishina Center, Hirosawa 2-1, Wako, Saitama 351-0198, Japan}
\author{N.~Kalantar-Nayestanaki}
\affiliation{KVI-CART, University of Groningen, Zernikelaan 25, 9747 AA Groningen, The Netherlands}
\author{S.~Kawase}
\affiliation{Department of Advanced Energy Engineering Science, Kyushu University, Kasuga, Fukuoka, 816-8580 Japan}
\author{S.~Kim}
\affiliation{Department of Physics and Astronomy, Seoul National University, 1 Gwanak-ro, Gwanak-gu, Seoul 08826, Republic of Korea}
\author{K.~Kisamori}
\affiliation{RIKEN Nishina Center, Hirosawa 2-1, Wako, Saitama 351-0198, Japan}
\author{T.~Kobayashi}
\affiliation{Department of Physics, Tohoku University, Miyagi 980-8578, Japan}
\author{D.~K\"orper}
\affiliation{GSI Helmholtzzentrum f\"ur Schwerionenforschung, 64291 Darmstadt, Germany}
\author{S.~Koyama}
\affiliation{Unversity of Tokyo, Tokyo 1130033, Japan}
\author{I.~Kuti}
\affiliation{Institute of Nuclear Research, Atomki, 4001 Debrecen, Hungary}
\author{V.~Lapoux}
\affiliation{Irfu, CEA, Universit\'e Paris-Saclay, 91191 Gif-sur-Yvette, France}
\author{S.~Lindberg}
\affiliation{Institutionen f\"or Fysik, Chalmers Tekniska H\"ogskola, 412 96 G\"oteborg, Sweden}
\author{S.~Masuoka}
\affiliation{Center for Nuclear Study, University of Tokyo, 2-1 Hirosawa, Wako, Saitama 351-0198, Japan}
\author{J.~Mayer}
\affiliation{Institut f\"ur Kernphysik, Universit\"at zu K\"oln, 50937 K\"oln, Germany}
\author{K.~Miki}
\affiliation{National Superconducting Cyclotron Laboratory, Michigan State University, East Lansing, Michigan 48824, USA}
\author{T.~Murakami}
\affiliation{Department of Physics, Kyoto University, Kyoto 606-8502, Japan}
\author{M.~Najafi}
\affiliation{KVI-CART, University of Groningen, Zernikelaan 25, 9747 AA Groningen, The Netherlands}
\author{K.~Nakano}
\affiliation{Department of Advanced Energy Engineering Science, Kyushu University, Kasuga, Fukuoka, 816-8580 Japan}
\author{N.~Nakatsuka}
\affiliation{Department of Physics, Kyoto University, Kyoto 606-8502, Japan}
\author{T.~Nilsson}
\affiliation{Institutionen f\"or Fysik, Chalmers Tekniska H\"ogskola, 412 96 G\"oteborg, Sweden}
\author{A.~Obertelli}
\affiliation{Irfu, CEA, Universit\'e Paris-Saclay, 91191 Gif-sur-Yvette, France}
\author{F.~de Oliveira Santos} 
\affiliation{Grand Acc\'el\'erateur National d'Ions Lourds (GANIL),
CEA/DRF-CNRS/IN2P3, Bvd Henri Becquerel, 14076 Caen, France}
\author{H.~Otsu}
\affiliation{RIKEN Nishina Center, Hirosawa 2-1, Wako, Saitama 351-0198, Japan}
\author{T.~Ozaki}
\affiliation{Department of Physics, Tokyo Institute of Technology, 2-12-1 O-Okayama, Meguro, Tokyo 152-8551, Japan}
\author{V.~Panin}
\affiliation{RIKEN Nishina Center, Hirosawa 2-1, Wako, Saitama 351-0198, Japan}
\author{S.~Paschalis}
\affiliation{Institut f\"ur Kernphysik, Technische Universit\"at Darmstadt, 64289 Darmstadt, Germany}
\author{D.~Rossi}
\affiliation{Institut f\"ur Kernphysik, Technische Universit\"at Darmstadt, 64289 Darmstadt, Germany}
\author{A.T.~Saito}
\affiliation{Department of Physics, Tokyo Institute of Technology, 2-12-1 O-Okayama, Meguro, Tokyo 152-8551, Japan}
\author{T.~Saito}
\affiliation{Unversity of Tokyo, Tokyo 1130033, Japan}
\author{M.~Sasano}
\affiliation{RIKEN Nishina Center, Hirosawa 2-1, Wako, Saitama 351-0198, Japan}
\author{H.~Sato}
\affiliation{RIKEN Nishina Center, Hirosawa 2-1, Wako, Saitama 351-0198, Japan}
\author{Y.~Satou}
\affiliation{Department of Physics and Astronomy, Seoul National University, 1 Gwanak-ro, Gwanak-gu, Seoul 08826, Republic of Korea}
\author{H.~Scheit}
\affiliation{Institut f\"ur Kernphysik, Technische Universit\"at Darmstadt, 64289 Darmstadt, Germany}
\author{F.~Schindler}
\affiliation{Institut f\"ur Kernphysik, Technische Universit\"at Darmstadt, 64289 Darmstadt, Germany}
\author{P.~Schrock}
\affiliation{Center for Nuclear Study, University of Tokyo, 2-1 Hirosawa, Wako, Saitama 351-0198, Japan}
\author{M.~Shikata}
\affiliation{Department of Physics, Tokyo Institute of Technology, 2-12-1 O-Okayama, Meguro, Tokyo 152-8551, Japan}
\author{Y.~Shimizu}
\affiliation{RIKEN Nishina Center, Hirosawa 2-1, Wako, Saitama 351-0198, Japan}
\author{H.~Simon}
\affiliation{GSI Helmholtzzentrum f\"ur Schwerionenforschung, 64291 Darmstadt, Germany}
\author{D.~Sohler}
\affiliation{Institute of Nuclear Research, Atomki, 4001 Debrecen, Hungary}
\author{L.~Stuhl}
\affiliation{RIKEN Nishina Center, Hirosawa 2-1, Wako, Saitama 351-0198, Japan}
\author{S.~Takeuchi}
\affiliation{Department of Physics, Tokyo Institute of Technology, 2-12-1 O-Okayama, Meguro, Tokyo 152-8551, Japan}
\author{M.~Tanaka}
\affiliation{Department of Physics, Osaka University, Osaka 560-0043, Japan}
\author{M.~Thoennessen}
\affiliation{National Superconducting Cyclotron Laboratory, Michigan State University, East Lansing, Michigan 48824, USA}
\author{H.~T\"ornqvist}
\affiliation{Institut f\"ur Kernphysik, Technische Universit\"at Darmstadt, 64289 Darmstadt, Germany}
\author{Y.~Togano}
\affiliation{Department of Physics, Tokyo Institute of Technology, 2-12-1 O-Okayama, Meguro, Tokyo 152-8551, Japan}
\author{T.~Tomai}
\affiliation{Department of Physics, Tokyo Institute of Technology, 2-12-1 O-Okayama, Meguro, Tokyo 152-8551, Japan}
\author{J.~Tscheuschner}
\affiliation{Institut f\"ur Kernphysik, Technische Universit\"at Darmstadt, 64289 Darmstadt, Germany}
\author{J.~Tsubota}
\affiliation{Department of Physics, Tokyo Institute of Technology, 2-12-1 O-Okayama, Meguro, Tokyo 152-8551, Japan}
\author{T.~Uesaka}
\affiliation{RIKEN Nishina Center, Hirosawa 2-1, Wako, Saitama 351-0198, Japan}
\author{Z.~Yang}
\affiliation{RIKEN Nishina Center, Hirosawa 2-1, Wako, Saitama 351-0198, Japan}
\author{M.~Yasuda}
\affiliation{Department of Physics, Tokyo Institute of Technology, 2-12-1 O-Okayama, Meguro, Tokyo 152-8551, Japan}
\author{K.~Yoneda}
\affiliation{RIKEN Nishina Center, Hirosawa 2-1, Wako, Saitama 351-0198, Japan}

\collaboration{SAMURAI21 collaboration}

\begin{abstract}
 Detailed spectroscopy of the neutron-unbound nucleus $^{28}$F has been performed for the first time following proton/neutron removal from $^{29}$Ne/$^{29}$F beams at energies around 230~MeV/nucleon. The invariant-mass spectra were reconstructed for both the $^{27}$F$^{(*)}+n$ and $^{26}$F$^{(*)}+2n$ coincidences and revealed a series of well-defined resonances. A near-threshold state was observed in both reactions and is identified as the $^{28}$F ground state, with $S_n(^{28}$F$)=-199(6)$~keV, while analysis of the $2n$ decay channel allowed a considerably improved $S_n(^{27}$F$)=1620(60)$~keV to be deduced.
 Comparison with shell-model predictions and eikonal-model reaction calculations have allowed 
spin-parity assignments to be proposed for some of the lower-lying levels of $^{28}$F.
 Importantly, in the case of the ground state, the reconstructed $^{27}$F$+n$ momentum distribution following neutron removal from $^{29}$F indicates that it arises mainly from the $1p_{3/2}$ neutron intruder configuration. This demonstrates that the island of inversion {around $N=20$} includes $^{28}$F, {and most probably $^{29}$F}, and suggests that $^{28}$O is not doubly magic.
\end{abstract}

\date{\today}
\maketitle

\header{Introduction}
 The study of nuclei located at the neutron dripline, beyond which they are no longer bound with respect to neutron emission, has become possible due to significant technological developments in high-intensity neutron-rich beams and high-efficiency detection arrays \cite{Nakamura2017}. Despite these advances, the neutron dripline is only accessible experimentally for light nuclei ($Z\lesssim10$) \cite{Ahn2019}, and even in this region it remains a theoretical challenge to predict it \cite{Erle2012}. Models incorporating the effect of three-nucleon forces \cite{Otsu2010,Barb2013,Ekst2013} have led to a better reproduction of the dripline.
 However, the effect of the continuum, which can drastically change the shell structure \cite{ Mich04,Kay17}, is not taken into account except for lighter nuclei \cite{Calc16}. The comparison between the isotopic chains of carbon, nitrogen and oxygen on the one hand, and of fluorine on the other, is particularly interesting: the dripline is located at $N=16$ for the former, while the fluorine chain extends to $N=22$ ($^{31}$F \cite{Ahn2019}). The reason for this, however, is not fully understood.

 In the fluorine chain, the odd neutron-number $^{28,30}$F isotopes are unbound, as they lack the extra binding energy provided by pairing. Christian et al.\ \cite{Chri2012} found that $^{28}$F is unbound by 220(50)~keV, and based on the agreement with the predictions of USDA/USDB shell-model calculations $^{28}$F was placed outside the {``Island of Inversion'' (IoI) \cite{IoI}}.
 This means that the ground state of $^{28}$F could be described by a particle-hole configuration ($\pi 0d_{5/2} \times \nu 0d_{3/2}^{-1}$) with respect to an unbound core of $^{28}$O, forming a multiplet of $J^{\pi}=1^+$--$\,4^+$ states as in $^{26}$F \cite{Lepa2013,Vande2017}.
 On the other hand, the relatively low energies of the first excited states in $^{27,29}$F suggest the presence of intruder neutron $pf$-shell contributions \cite{Door2017}. If $^{28}$F contains such contributions, negative-parity states, like the $J^{\pi}=1^-$--$\,4^-$ multiplet resulting from the $\pi 0d_{5/2}\otimes \nu 1p_{3/2}$ coupling, should appear at low energy.

 The location of the dripline in fluorine at $N=22$ suggests a profound change in shell structure around doubly-magic $^{24}$O \cite{Hoff2009,Kanu2009,Tsho2012}. A direct experimental signature of these structural changes can be found in the evolution of the energies of the $3/2^+$, $7/2^-$ and $3/2^-$ states, arising from the neutron $0d_{3/2}$, $0f_{7/2}$ and $1p_{3/2}$ orbits, in the $N=17$ isotones from $Z=14$ to 10 (Fig.~3 of \cite{Sorl2013}).
 In $^{31}_{14}$Si, the spacing between the ground $3/2^+$ and the $7/2^-$ states, which is linked to the size of the $N=20$ gap, is 3.2~MeV, and the $3/2^-$ state lies 0.5~MeV above the $7/2^-$.
 In $^{27}_{10}$Ne, the $3/2^+$--$7/2^-$ gap is reduced to 2~MeV, and the $3/2^-$ level moves below the $7/2^-$ state, at only 0.8~MeV above the $3/2^+$ g.s.\ \cite{Terr2006, Ober2006,Brow2009}.
 In $^{29}$Ne, the $3/2^-$ intruder state becomes the ground state \cite{Koba2016}. 
 This migration of levels has been suggested to be due to the hierarchy of the $p$-$n$ forces present above $^{24}$O \cite{Sorl2013}, and in particular to the central and tensor components \cite{Utsu1999,Otsu2005,Smir2012}. 
           

 {This Letter reports on the first detailed spectroscopic study of $^{28}$F, which has been carried out using proton and neutron removal from high-energy $^{29}$Ne and $^{29}$F beams, respectively.
 In the former reaction, the $^{29}$Ne neutron configuration will remain unchanged and negative-parity states are expected to be populated at low energy in $^{28}$F through the removal of a $0d_{5/2}$ proton.
 Neutron removal, however, can lead to both positive- and negative-parity levels in $^{28}$F depending on the degree to which intruder ($2p2h$ and beyond) configurations are present in $^{29}$F.
 This study was possible owing to the high luminosity provided by a thick liquid H$_2$ target and the relatively intense secondary beams, coupled with state-of-the-art arrays for the detection of the high-energy neutrons and charged fragments and of the de-excitation $\gamma$-rays.
 The results indicate that $^{28}$F, and most probably $^{29}$F, lie within the IoI, and also suggest that $^{28}$O is not doubly magic.}

\begin{figure*}[t]
  \includegraphics[width=\textwidth]{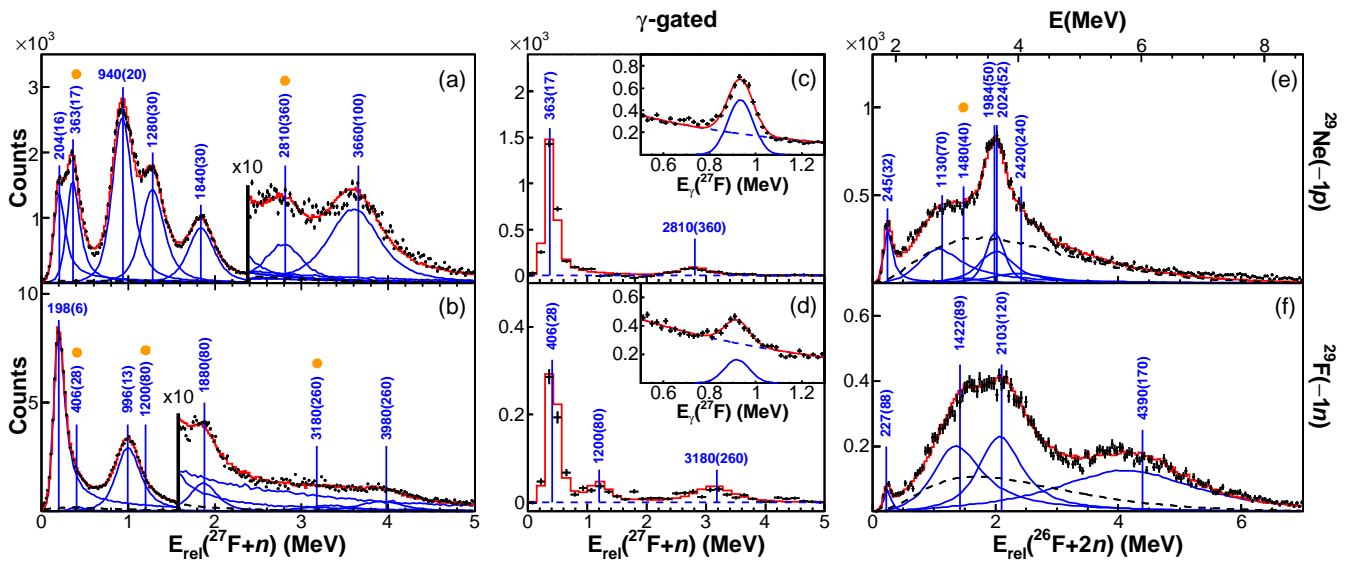}
  \caption{{Left: relative-energy spectra of the $^{27}$F$+n$ system populated from the reactions (a) $^{29}$Ne($-1p$) and (b) $^{29}$F($-1n$). Right: same for the $^{26}$F$+2n$ system populated from (e) $^{29}$Ne($-1p$) and (f) $^{29}$F($-1n$). The fit in red corresponds to a sum of resonances (blue, with the resonance energy in keV) plus a non-resonant distribution (dashed black).
 Center: same as left, obtained in coincidence with the 915~keV excited state of $^{27}$F (after background subtraction) populated from (c) $^{29}$Ne($-1p$) and (d) $^{29}$F($-1n$).
 The energy axis E on the top right is given with respect to $S_n$($^{28}$F), and orange dots mark resonances in coincidence with the corresponding fragment $\gamma$-rays (see Fig.~\ref{lvl_scheme}).}}
  \label{Ed} 
\end{figure*}

\header{Experimental setup}
 The experiment was performed at the Radioactive Isotope Beam Factory (RIBF) of the RIKEN Nishina Center. Secondary beams of $^{29}$Ne ($\sim8.1$~kHz, 228~MeV/nucleon) and $^{29}$F ($\sim90$~Hz, 235~MeV/nucleon) were produced by fragmentation of a 345~MeV/nucleon $^{48}$Ca beam ($\sim500$~pnA) on a 15mm-thick Be target, and prepared using the BigRIPS fragment separator \cite{Kubo2003,Ohnishi2010}.
 Secondary beam particles were identified via their energy loss and time of flight as measured using thin plastic scintillators, and tracked on to the object point of the SAMURAI spectrometer \cite{Kobayashi2013} using two sets of multi-wire drift chambers, where the \mbox{MINOS} target \cite{Obertelli2014} was located. The latter consisted of a 15cm-thick liquid-hydrogen cell surrounded by a time-projection chamber, that allowed the reconstruction of the reaction vertex with a precision of 3~mm (sigma) in the beam direction using the intersection between the trajectory of the incoming beam and the measured track(s) of the outgoing proton(s) for the $(p,pn)$ and $(p,2p)$ reactions.
 The DALI2 NaI array \cite{Takeuchi2014} surrounded the target for the detection of the in-flight de-excitation of fragments (with an efficiency of $\varepsilon_\gamma\sim15\%$ at 1~MeV). 

 The beam-velocity reaction products were detected in the forward direction using the SAMURAI setup, including the NEBULA \cite{Kondo15} and NeuLAND demonstrator \cite{NeuLAND} neutron arrays, placed respectively some 14 and 11~m downstream of the target.
 The \mbox{SAMURAI} superconducting dipole magnet \cite{Sato-SAMURAI}, with a central field of 2.9~T and a vacuum chamber equipped with thin exit windows \cite{Shimizu}, provided for the momentum analysis of the charged fragments.
 Their trajectories and magnetic rigidity were determined using drift chambers at the entrance and exit of the magnet \cite{Kobayashi2013}. This information, combined with the energy loss and time of flight measured using a 24-element plastic scintillator hodoscope, provided the identification of the projectile-like fragments.
 The neutron momenta were derived from the time of flight, with respect to a thin plastic start detector positioned just upstream of the target, and the hit position as measured using the NEBULA and NeuLAND arrays \cite{Kondo2020}, with efficiencies of $\varepsilon_n\sim50\%$ and $\varepsilon_{nn}\sim10\%$ for decay energies of 0--3~MeV.

\header{Energy spectra}
 The relative energy ($E_{rel}$) of the unbound $^{28}$F system was reconstructed from the momenta of the $^{26,27}$F fragments and neutron(s) \cite{Kondo2020}. 
 The $^{27}$F+$n$ spectra {for both reactions are shown on the left of} Fig.~\ref{Ed}. The resolution is considerably improved compared to previous studies of neutron-unbound systems \cite{Chri2012,Kondo2016}, {owing to the high-granularity NeuLAND array as well as the MINOS target. The resolution of $E_{rel}(^{27}$F+$n$) varied as $\mbox{\sc{fwhm}}\sim0.18\,E_{rel}^{\,0.63}$~MeV}.
 In order to deduce the character of resonances in $^{28}$F, the spectra were described using single-level R-matrix line-shapes \cite{BW}, which were used as the input for a complete simulation of the setup (including the beam characteristics, the reaction, and the detector resolutions and acceptances), combined with a non-resonant component obtained from event-mixing \cite{Giacomo14,Leblond18} and from the simulation of independent fragments and neutrons, respectively for the two- and three-body spectra. The results of the fit are listed on the figure and summarized in Ref.~\cite{SuppMatTables}.

\begin{figure}[t]
 \hspace*{-4mm}\includegraphics[width=9cm,clip=]{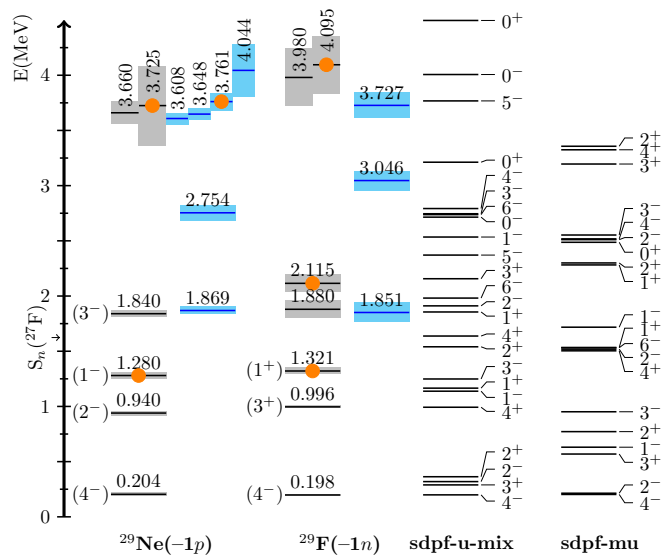}
  \caption{Energies of the resonances observed in $^{28}$F in the $1n$ (black) and $2n$ (blue) decay channels compared to shell-model calculations (the g.s.\ energies are normalized to experiment). The resonances observed in the $2n$ channel are placed in the level scheme according to {$S_n(^{27}$F$)=1.620(60)$~MeV (see text).
 The grey and blue bands represent the uncertainty in the resonance energies. Levels with an orange dot decay to excited states of $^{27}$F ($1n$ decay) or $^{26}$F ($2n$ decay)}.}
  \label{lvl_scheme} 
\end{figure}

 The energy spectra of Fig.~\ref{Ed}(a,b), from the $^{29}$Ne$(-1p)$ and $^{29}$F$(-1n)$ reactions, {exhibit a lowest-lying resonance with a width of $\Gamma=180(40)$~keV at respectively 204(16) and 198(6)~keV above threshold, without any coincident $\gamma$-ray.
 The weighted mean, $199(6)$~keV, provides therefore a determination of the g.s.\ energy of $^{28}$F ($-S_n$)}.
 This is compatible with the less precise value of 220(50)~keV from Ref.~\cite{Chri2012} using the $^{29}$Ne$(-1p)$ reaction. 
 As shown in Fig.~\ref{Ed}(a), we observe a second peak in the $(-1p)$ channel at 363(17)~keV, which is in coincidence with the 915(12)~keV transition (inset of Fig.~\ref{Ed}c) from the decay of the excited state of $^{27}$F \cite{Door2017}.
 {As such, the resonance lies at the sum energy of 1278(21)~keV above threshold}.
 As this value matches the energy of the fourth peak at 1280(30)~keV, we propose that the 1280~keV state, populated in $^{29}$Ne$(-1p)$, decays both to the ground and first-excited states of $^{27}$F, with corresponding branching ratios of 60\% and 40\%.
 The 2810~keV resonance is also observed in coincidence with the 915~keV $\gamma$-ray. It is thus placed at an energy of 3725~keV (Fig.~\ref{lvl_scheme}). Three other resonances identified in Fig.~\ref{Ed}(a) at 940, 1840 and 3660~keV are also placed in Fig.~\ref{lvl_scheme}.

 The spectrum of Fig.~\ref{Ed}(b), obtained from $^{29}$F$(-1n)$, displays three clear resonances, including the g.s.\ (see above). The resonance at 996(13)~keV does not fully match the 940(20)~keV observed in the $(-1p)$ reaction. We thus propose that they correspond to two different states, as shown in Fig.~\ref{lvl_scheme}. {Given the uncertainties, the 1880(80) and 1840(30)~keV resonances observed in both reactions can correspond to the same state}.
 If we require a coincidence with the 915~keV $\gamma$-ray of $^{27}$F, one can see in Fig.~\ref{Ed}(d) the two resonances at 406(28) and 3180(260)~keV plus an additional structure at 1200(80)~keV, corresponding therefore respectively to levels at 1321, 4095 and 2115~keV (Fig.~\ref{lvl_scheme}).
 Note that the 406(28)~keV resonance overlaps with that at 363(17)~keV, which was proposed to decay in competition with the 1280(30)~keV transition in the $(-1p)$ channel with similar intensities. However, as the fit of the $(-1n)$ data does not allow the placement of a 1280(30)~keV resonance with the required intensity, we propose that the 363(17) and 406(28)~keV transitions come from the decay of different states located respectively at 1280 and 1321~keV. Finally a resonance is placed at 3980~keV.

 Resonances in $^{28}$F decaying by $2n$ emission have been identified after applying cross-talk rejection conditions to the $^{26}$F$+2n$ events \cite{crosstalk}. As can be seen in Fig.~\ref{Ed}(e,f), the lowest-lying peak produced in both the $(-1p)$ and $(-1n)$ reactions has compatible energies of respectively $E_{rel}=245(32)$ and 227(88)~keV.
 The states observed in the $2n$ decay correspond to excitation energies of $E_{rel}+S_{n}(^{27}$F), when referenced to the $^{28}$F g.s., or to an excitation energy of $E_{rel}+S_{2n}(^{28}$F).
 According to AME2016 \cite{Huang2017}, the uncertainty on $S_{n}(^{27}$F$)=1270(410)$~keV is large, which also influences the present determination of $S_{2n}(^{28}$F), making the placement of the resonances very uncertain.

 However, we first note that the two low-energy resonances are, as for the 1840 and 1880~keV resonances in the $^{27}$F$+n$ decay, produced in both reactions. Second, they have compatible intrinsic widths \cite{SuppMatTables}, independent of the decay mode. Third, the ratios between the 245 and 1840~keV resonances in $(-1p)$, and the 227 and 1880~keV resonances in $(-1n)$, are the same ($\sim10\%$). This suggests that they all originate from a single state at $\sim1860$~keV, that decays both by $1n$ and $2n$ emission. Excellent agreement between the $1n$ and $2n$ decay spectra is obtained using $S_n(^{27}$F$)=1620(60)$~keV and $S_{2n}(^{28}$F$)=1420(60)$~keV, the latter being deduced from the present determination of $S_{n}(^{28}$F).
 A summary of all the levels identified is reported in Fig.~\ref{lvl_scheme}. 

\begin{figure}[t]
 \includegraphics[height=3.5cm]{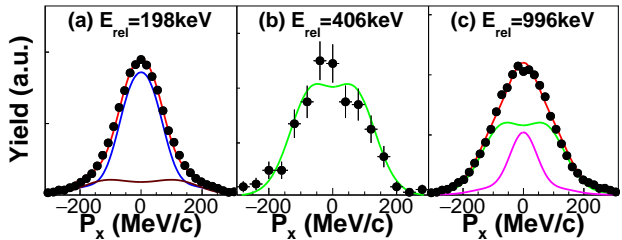}
  \caption{Experimental transverse-momentum distributions 
of the $^{27}$F$+n$ system following the $^{29}$F$(-1n)$ reaction compared to eikonal-model calculations for:
 (a) removal of a neutron with $\ell=1,3$ (respectively blue, brown lines) for the g.s.\ at $E_{rel}=198$~keV; (b) a pure $\ell=2$ distribution for the resonance at 406~keV (corresponding to the state at 1321~keV); (c) a mixture of $\ell=0,2$ (respectively purple, green lines) for the state at $E_{rel}=996$~keV. In (a,c) the red line is the total fit.}
  \label{p_dist} 
\end{figure}

\header{Momentum distributions}
 In the $(-1n)$ reaction, the reconstructed momentum distribution of the $^{27}$F$+n$ system allows the orbital angular momentum of the removed neutron to be deduced \cite{Hansen2003}.
 The transverse-momentum distribution corresponding to the feeding of the $^{28}$F g.s.\ is fitted in Fig.~\ref{p_dist}(a) with eikonal-model calculations \cite{Tos01,SuppMatEikonal} using a combination of $\ell=1,3$ components.
 This choice of negative-parity $\ell$ values is guided by the fact that the g.s.\ is also produced in the $^{29}$Ne$(-1p)$ reaction, which, as discussed earlier, is expected to lead to negative-parity states at low $E_{rel}$. The fit, which gives a spectroscopic factor of $C^2S=0.40(6)$, is dominated by the $\ell=1$ component ($79\%$), meaning that the g.s.\ of $^{28}$F 
 is mainly composed of an intruder $p$-wave component. 

 The momentum distribution of the resonance at 406~keV, Fig.~\ref{p_dist}(b), is obtained after gating on the 915~keV $\gamma$-ray transition. It is well reproduced by a pure $\ell=2$ component, meaning that the parity of the 1321~keV state is positive, with $C^2S=0.012(4)$. In order to account for its highly favored $1n$ decay through the $(1/2^+)$ excited state of $^{27}$F, rather than to the $(5/2^+)$ g.s.\ despite the higher energy available, we propose that it has $J^{\pi}=1^+$. Indeed, this would result in an $\ell=0$ neutron decay to the excited state, as opposed to an $\ell=2$ decay to the ground state. Other (higher) spin values would not account for such a unique behavior.  
 For the resonance at 996~keV, the momentum distribution, Fig.~\ref{p_dist}(c), is very well reproduced by an admixture of $\ell=2$ ($72\%$) and $\ell=0$ ($28\%$), making it another candidate for a positive-parity state, with $C^2S=0.30(4)$.

 As for the $(-1p)$ reaction, the four most populated states, with energies 204, 940, 1280 and 1840~keV, all display momentum distributions compatible with $\ell=2$ proton removal from the $d_{5/2}$ orbital, with $C^2S$ of respectively 0.20(3), 0.46(7), 0.50(8) and 0.22(4), summing up to about 1.4, as compared to the maximal expected occupancy of 2 for the $d_{5/2}$ orbital in $^{29}$Ne.

\header{Shell-model calculations}
 These have been performed using the {\it sdpf-u-mix} interaction \cite{sdpfmix} in order to predict the energy, $J^\pi$ (Fig.~\ref{lvl_scheme}, right) and $C^2S$ values of negative- and positive-parity states in $^{28}$F. In order to assess the sensitivity to the level scheme, the {\it sdpf-mu} interaction \cite{Utsuno2012} has also been used.
 The {\it sdpf-u-mix} interaction has been refined in order to reproduce the observed $3/2^-$ and $7/2^-$ level crossing and location of the $pf$ intruder orbitals in $^{27}$Ne, $^{29}$Mg and $^{31}$Si, and the dripline at $^{31}$F. 

 Both calculations predict about 15 negative- and positive-parity states below 2~MeV, demonstrating that the normal and intruder configurations in $^{28}$F are very close in energy. The first 10 states have relatively pure configurations (60--80\%) mostly originating from the proton $0d_{5/2}$ and neutron $0d_{3/2}$ and $1p_{3/2}$ orbits, with the exception of the $5^-$ and $6^-$ levels that arise from a neutron in the $0f_{7/2}$ orbit. The $\pi 0d_{5/2} \otimes \nu 1p_{3/2}$  and $\pi 0d_{5/2} \otimes \nu 0d_{3/2}$  couplings lead to a multiplet of $J=1$--4 states with negative and positive parity, respectively. 

 The calculations predict that four negative-parity states $J^\pi=(4,2,1,3)^-$ are mainly populated in the $(-1p)$ reaction with dominant $\ell=1$ components and $C^2S$ values of 0.75, 0.44, 0.35 and 0.19, in rather good agreement with experiment. {We thus think we have populated this multiplet of states. Among them,} a $J^\pi=4^-$ g.s.\ is predicted by both calculations, with $\Gamma$ of about 180~keV, in agreement with experiment.
 Using similar arguments, the 940~keV state is proposed to be $J^\pi=2^-$. The $1^-$ level is predicted to decay both to the ground ($5/2^+$) and first-excited ($1/2^+$) states of $^{27}$F with $\ell=1$, and could correspond to the state identified at 1280~keV. As it has the highest energy in both calculations, the 1840-keV resonance is tentatively assigned as $J^\pi=3^-$.

 In the $(-1n)$ reaction, the $4^-$ g.s.\ is calculated to be the most populated among other negative-parity states with $C^2S=0.36$, coming mostly (90\%) from an $\ell=1$ removal, to be compared with $C^2S=0.40(6)$, with 79\% of $\ell=1$ fraction.
 As for the positive-parity states, produced only in the $(-1n)$ reaction, both the  {\it sdpf-u-mix} interaction predicts the lowest state as $J^{\pi}=3^+$ with $C^2S=0.54$, in reasonable agreement with the 996~keV state with $C^2S=0.30(4)$. The $1^+$ state is predicted to decay principally to the first excited state of $^{27}$F with $\ell=0$, making the 1321~keV state a good $J^\pi=1^+$ candidate. The calculated $C^2S$ value of the $1^+$ state, 0.31, is however much larger than experiment.  

 The first positive-parity states are predicted too low in energy, which could be explained by effects of the continuum (not taken into account explicitly in the present calculations) that change the effective two-body matrix elements \cite{Vande2017,Stef14} and induce lingering of the $\ell=1$ states compared to $\ell=2$ \cite{Kay17}. Another feature that could be related to the effects of the continuum, discussed in Ref.~\cite{Mich04} as an apparent reduction of pairing, is the damping of the $|S_n(N)-S_n(N+1)|$ amplitude when approaching the dripline. While these amplitudes are correctly reproduced in lighter ($N\leqslant16$) fluorine isotopes by the present calculations, our experimental $S_n(^{27}$F$)-S_n(^{28}$F) value of 1.82(6)~MeV is significantly smaller than the predicted 2.8~MeV.

\header{Conclusions}
 In summary, {detailed spectroscopy of $^{28}$F has been undertaken using nucleon removal from secondary beams of $^{29}$F and $^{29}$Ne, with statistics orders of magnitude higher than the previous study and unprecedented energy resolution.
 This was made possible through the unique combination of a thick liquid target and state-of-the-art arrays for the detection of high-energy neutrons and charged fragments, as well as de-excitation $\gamma$-rays. They proved essential to cope with the high density of states in $^{28}$F and allowed the identification of the $1n$ and $2n$ decay modes, including transitions to bound excited states of $^{26,27}$F.
 In addition to making comparisons with shell-model calculations, the $^{28}$F transverse-momentum distributions following neutron removal, combined with eikonal-model calculations, allowed the $\ell$ configuration of the removed neutron to be deduced}.
 
 The $^{28}$F g.s.\ resonance was unambiguously identified, with $S_n(^{28}$F$)=-199(6)$~keV. It has a negative parity with an $\ell=1$ content of about 80\%, which places $^{28}$F inside the IoI. Based on the comparison to shell-model calculations of the decay patterns, resonance widths and $C^2S$ values, we propose that the multiplet of $J^\pi=(1$--$4)^-$ states originating from the $\pi 0d_{5/2} \otimes \nu 1p_{3/2}$ configuration 
 has been identified.
 The first positive-parity resonance ($3^+$) is proposed at 996~keV, about 560~keV higher than shell-model predictions. A candidate for a $J^\pi=1^+$ resonance is proposed at 1321~keV. As opposed to $^{26}$F, that has well-identified positive-parity states from $p$-$n$ configurations above a doubly-magic $^{24}$O core, $^{28}$F displays mixed negative- and positive-parity states, with the negative-parity states being more bound. These features strongly suggest that {$N=20$} magicity is not restored at $^{28}$O.
 {Moreover, the single-neutron removal, including the strong $\ell=1$ feeding of the negative-parity $^{28}$F g.s., supports the suggestion, based on mass measurements, that $^{29}$F also lies within the IoI \cite{Gaud12}}.

 Finally, we propose a very precise value of $S_n(^{27}$F$)=1620(60)$~keV, as compared to the tabulated value of $1270(410)$~keV, which combined with $S_n(^{28}$F$)=-199(6)$~keV leads to a reduced oscillation in the $S_n$ values of about 35\% at the dripline, as compared to shell-model calculations. This damping in the oscillations has also been recently observed in the boron isotopic chain \cite{Leblond18}, suggesting that a reduced pairing force may be a generic feature of dripline nuclei.

\acknowledgments

 We thank M.~Ploszajczak for fruitful discussions, and the accelerator staff of the RIKEN Nishina Center for their efforts in delivering the intense $^{48}$Ca beam.
 N.L.A., F.D., J.G., F.M.M.\ and N.A.O.\ acknowledge partial support from the Franco-Japanese LIA-International Associated Laboratory for Nuclear Structure Problems as well as the French ANR-14-CE33-0022-02 EXPAND.
 J.A.T.\ acknowledges support from the Science and Technology Facilities Council (U.K.) grant No.\ ST/L005743/1. I.G.\ was supported by HIC for FAIR and Croatian Science Foundation under projects No.\ 1257 and 7194. Z.D., Z.E.\ and D.S.\ were supported by projects No.\ GINOP-2.3.3-15-2016-00034 and K128947, and I.K.\ by project No.\ PD 124717. J.K.\ acknowledges support from RIKEN as short-term International Program Associate. This material is based upon work supported by the U.S.\ Department of Energy, Office of Science, Office of Nuclear Physics, under contract No.\ DE-AC02-06CH11357 (ANL). {This project was funded in part by the Deutsche Forschungsgemeinschaft (DFG, German Research Foundation), Project-ID\ 279384907, SFB\ 1245, and the GSI-TU Darmstadt cooperation agreement.}


\begin{thebibliography}{99}

\bibitem{Nakamura2017}	T.~Nakamura \etal, Prog.\ Part.\ Nucl.\ Phys.\ {\bf97}, 53 (2017).
\bibitem{Ahn2019}		D.S.~Ahn \etal, Phys.\ Rev.\ Lett.\ {\bf123}, 212501 (2019).
\bibitem{Erle2012}	J.~Erler \etal, Nature (London) {\bf486}, 509 (2012).
\bibitem{Otsu2010}	T.~Otsuka \etal, Phys.\ Rev.\ Lett.\ {\bf105}, 032501 (2010).
\bibitem{Barb2013}	A.~Cipollone, C.~Barbieri and P. Navr\'atil, Phys.\ Rev.\ Lett.\ {\bf111}, 062501 (2013).
\bibitem{Ekst2013}	A.~Ekstr\"om \etal, Phys.\ Rev.\ Lett.\ {\bf110}, 192502 (2013).
\bibitem{Mich04}	N.~Michel \etal, Acta Physica Polonica B35, 1249 (2004).
\bibitem{Kay17}		B.P.~Kay, C.R.~Hoffman and A.O.~Machiavelli, Phys.\ Rev.\ Lett.\ {\bf119}, 182502 (2017).
\bibitem{Calc16}	A.~Calci \etal, Phys.\ Rev.\ Lett.\ {\bf117}, 242501 (2016).
\bibitem{Chri2012}	G.~Christian \etal, Phys.\ Rev.\ Lett.\ {\bf108}, 032501 (2012).
\bibitem{IoI}		W.~Warburton, J.A.~Becker and B.A.~Brown, Phys.\ Rev.\ C {\bf41}, 1147 (1990).
\bibitem{Lepa2013}	A.~Lepailleur \etal, Phys.\ Rev.\ Lett.\ {\bf110}, 082502 (2013).
\bibitem{Vande2017}	M.~Vandebrouck \etal, Phys.\ Rev.\ C {\bf96}, 054305 (2017).
\bibitem{Door2017}	P.~Doornenbal \etal, Phys.\ Rev.\ C {\bf95}, 041301 (2017). 
\bibitem{Hoff2009}	C.R.~Hoffman \etal, Phys.\ Lett.\ B {\bf672}, 17 (2009).
\bibitem{Kanu2009}	R.~Kanungo \etal, Phys.\ Rev.\ Lett.\ {\bf102}, 152501 (2009).
\bibitem{Tsho2012}	K.~Tshoo \etal, Phys.\ Rev.\ Lett.\ {\bf109}, 022501 (2012).

\bibitem{Sorl2013}	O.~Sorlin, Eur.\ Phys.\ J.\ Web of conferences 66, 01016 (2014).
\bibitem{Terr2006}	J.R.~Terry \etal, Phys.\ Lett.\ B {\bf640}, 186 (2006).
\bibitem{Ober2006}	A.~Obertelli \etal, Phys.\ Lett.\ B {\bf633}, 33 (2006).
\bibitem{Brow2009}	S.M.~Brown \etal, Phys.\ Rev.\ C {\bf79}, 0143010 (2009).

\bibitem{Koba2016}	N.~Kobayashi \etal, Phys.\ Rev.\ C {\bf93}, 014613 (2016).
\bibitem{Utsu1999}	Y.~Utsuno \etal, Phys.\ Rev.\ C {\bf60}, 054315 (1999).
\bibitem{Otsu2005}	T.~Otsuka \etal, Phys.\ Rev.\ Lett.\ {\bf95}, 232502 (2005).
\bibitem{Smir2012}	N.A.~Smirnova \etal, Phys.\ Rev.\ C {\bf86}, 034314 (2012).
  
\bibitem{Kubo2003}		T.~Kubo, Nucl.\ Instum.\ Meth.\ Phys.\ Res.\ B {\bf204}, 97 (2003).
\bibitem{Ohnishi2010}	T.~Ohnishi \etal, J.\ Phys.\ Soc.\ Jpn.\ {\bf79}, 073201 (2010).
\bibitem{Kobayashi2013}	T.~Kobayashi \etal, Nucl.\ Instum.\ Meth.\ Phys.\ Res.\ B {\bf317}, 294 (2013).
\bibitem{Obertelli2014}	A.~Obertelli \etal, Eur.\ Phys.\ J.\ A {\bf50}, 8 (2014).
\bibitem{Takeuchi2014}	S.~Takeuchi \etal, Nucl.\ Instum.\ Meth.\ Phys.\ Res.\ A {\bf763}, 596 (2014).
\bibitem{Kondo15}		T.~Nakamura and Y.~Kondo, Nucl.\ Instrum.\ Meth. Phys.\ Res.\ B {\bf376}, 1 (2015).
\bibitem{NeuLAND}	Technical report for the design, construction and commissioning of NeuLAND, https://edms.cern.ch/ui/file/1865739/2/TDR\_R3B\_NeuLAND\_public.pdf
\bibitem{Sato-SAMURAI}	H.~Sato \etal, IEEE Trans.\ Applied Superconductivity {\bf23}, 4500308 (2013).
\bibitem{Shimizu}	Y.~Shimizu \etal, Nucl.\ Instrum.\ Meth. Phys.\ Res. B {\bf317}, 739 (2013).
\bibitem{Kondo2020}	Y.~Kondo, T.~Tomai and T.~Nakamura, Nucl.\ Instrum.\ Meth. Phys.\ Res.\ B {\bf463}, 173 (2020).
\bibitem{Kondo2016}	Y.~Kondo \etal, Phys.\ Rev.\ Lett.\ {\bf116}, 102503 (2016).
  
\bibitem{BW}		A.~M.~Lane and R.G.~Thomas, Rev.\ Mod.\ Phys.\ {\bf30}, 257 (1958).
\bibitem{Giacomo14}	G.~Randisi \etal, Phys.\ Rev.\ C {\bf89}, 034320 (2014).
\bibitem{Leblond18}	S.~Leblond \etal, Phys.\ Rev.\ Lett.\ {\bf121}, 262502 (2018).
\bibitem{SuppMatTables}	See Supplemental Material \Red{[url to be added]} for the energies and widths obtained from the fits of $^{28}$F (Table~S1 and Table~S2).
\bibitem{crosstalk}	T.~Nakamura and Y.~Kondo, Nucl.\ Instrum.\ Meth. Phys.\ Res. B {\bf376}, 156 (2016).

\bibitem{Huang2017}	W.J.~Huang \etal, Chinese Physics C {\bf41}, 030002 (2017).
\bibitem{decayscheme}	See Supplemental Material \Red{[url to be added]} for the experimental decay schemes of $^{28}$F in Fig.~S1. 

\bibitem{Hansen2003}	P.G.~Hansen and J.A.~Tostevin, Direct Reactions With Exotic Nuclei, Annu.\ Rev.\ Nucl.\ Part.\ Sci.\ {\bf53}, 219 (2003).
\bibitem{Tos01}			J.A.~Tostevin, Nucl.\ Phys.\ A {\bf682}, 320 (2001).
\bibitem{SuppMatEikonal}	See Supplemental Material \Red{[url to be added]} for details on the proton target considerations in the direct nucleon removal model calculations, which include Refs.~\cite{AT,CG,Ray,abu,Bro98,Typ,Ric}.
\bibitem{AT}	J.S.~Al-Khalili and J.A.~Tostevin, Phys.\ Rev.\ C {\bf57}, 1846 (1998).
\bibitem{CG}	S.K.~Charagi and S.K.~Gupta, Phys.\ Rev.\ C {\bf41}, 1610 (1990).
\bibitem{Ray}	L.~Ray, Phys.\ Rev.\ C {\bf20}, 1857 (1979).
\bibitem{abu}	B.~Abu-Ibrahim \etal, Phys.\ Rev.\ C {\bf77}, 034607 (2008).
\bibitem{Bro98} B.A.~Brown, Phys.\ Rev.\ C {\bf58}, 220 (1998).
\bibitem{Typ}	B.A.~Brown, S.~Typel, and W.A.~Richter, Phys.\ Rev.\ C {\bf65}, 014612 (2002).
\bibitem{Ric}	W.A.~Richter and B.A.~Brown, Phys.\ Rev.\ C {\bf67}, 034317 (2003).
\bibitem{sdpfmix}	E.~Caurier, F.~Nowacki and A.~Poves, Phys.\ Rev.\ C {\bf90}, 014302 (2014).
\bibitem{Utsuno2012}Y. Utsuno \etal, Phys.\ Rev.\ C {\bf86}, 051301 (2012).
\bibitem{Stef14}	I. Stefan \etal, Phys.\ Rev.\ C {\bf90}, 014307 (2014).
\bibitem{Gaud12}	L.~Gaudefroy \etal, Phys.\ Rev.\ Lett.\ {\bf109}, 202503 (2012).
 
\end{thebibliography}
\end{document}